# Crowd-Sourcing Fuzzy and Faceted Classification for Concept Search

RICHARD ABSALOM, The Hague
DAP HARTMANN, Delft University of Technology
MARKUS LUCZAK-RÖSCH, University of Southampton
ASKE PLAAT, Tilburg University

## 1. INTRODUCTION

Searching for concepts in science and technology is often a difficult task. To facilitate concept search, different types of human-generated metadata have been created to define the content of scientific and technical disclosures. Classification schemes such as the International Patent Classification (IPC)[1] and MEDLINE's MeSH[2] are structured and controlled, but require trained experts and central management to restrict ambiguity (Mork, 2013). While unstructured tags of folksonomies can be processed to produce a degree of structure (Kalendar, 2010; Karampinas, 2012; Sarasua, 2012; Bragg, 2013) the freedom enjoyed by the crowd typically results in less precision (Stock 2007).

Existing classification schemes suffer from inflexibility and ambiguity. Since humans understand language, inference, implication, abstraction and hence concepts better than computers, we propose to harness the collective wisdom of the crowd. To do so, we propose a novel classification scheme that is sufficiently intuitive for the crowd to use, yet powerful enough to facilitate search by analogy, and flexible enough to deal with ambiguity. The system will enhance existing classification information. Linking up with the semantic web and computer intelligence, a Citizen Science effort (Good, 2013) would support innovation by improving the quality of granted patents, reducing duplicitous research, and stimulating problem-oriented solution design.

A prototype of our design is in preparation. A crowd-sourced fuzzy and faceted classification scheme will allow for better concept search and improved access to prior art in science and technology.

## 2. THE CONTEXT

To demonstrate our design we will consider the case of artificial sharkskin lining solving the problem of biliary stent occlusion. The artificial sharkskin lining resists adhesion of colloids and other particles (Donelli, 2007).

If such a lining were the subject matter of a patent application, a *prior art search* would include a search for similar disclosures that would render any difference obvious. Different implementations of the idea, expressed in different terminology, are poorly served by a keyword search in the text. The problem may be further compounded by the obfuscation of patent legalese. For example, a document disclosing a '*conduit with boundary layer turbulence-inducing non-smooth interior surface*' may be very pertinent, although this phrase does not explicitly mention either 'stent', 'sharkskin' or 'occlusion'.

Patent offices have traditionally approached prior art search through the development of human-implemented classification schemes which group similar documents in what were originally paper collections (Wolter, 2012; Wu, 2007). The IPC, and more recently the Cooperative Patent

---

[1] www.wipo.int/classifications/ipc/
[2] http://www.ncbi.nlm.nih.gov/pubmed/

Classification (CPC)[3] have so far been successful, although their paper-library origins are apparent. The structure of the scheme is root based, and, as a consequence, there are no inferences (links) between different technological areas. Smartphones, for example, may be classified in either computing or telecommunications (the latter applies in practice). This lack of commonality is not representative of modern-day technology and hampers effective search. Furthermore, the patent office classification scheme does not cover the vast volumes of academic non-patent literature, given that the schemes are not crowd compatible. Thus, it is effectively not feasible to search for all relevant prior art for a patent application.

A searcher with the problem of stent occlusion, but not aware of the sharkskin solution may trawl through the literature, including 'free' solutions disclosed in expired patents. Alternatively a method from a different technical area of technology could inspire a novel solution. Could not the biomimicry of artificial sharkskin on the hull of ships, used to reduce drag and prevent fouling, suggest a solution (Donelli, 2007; Bragg, 2013)? This cross-area type of search is what we call *solution search*.

3. THE DESIGN

We propose a new classification scheme using *faceted* classification to define concepts (Broughton, 2012). We combine 5 facets for (1) the technology/science (doing/knowing), in our example surface materials; (2) the application: notably the stent; (3) the operating mode: the use in-situ rather than the design, manufacture, test, insertion or re-cycling; (4) the problem: the occlusion; and (5) the solution: the simulated sharkskin lining.

Faceted classification in classical (Boolean) classification schemes (Mork 2013; Wolter, 2012; Bogers, 2006; Bonino, 2010) relies on the mutual exclusivity of the different classes within a facet. As a direct consequence, classification is exclusive and must be accurate. Boolean classification is a poor approximation to information, which is generally of a fuzzy nature.[4] Furthermore, as the prior art grows and ambiguity of information increases, so does the complexity of Boolean schemes, and the likelihood of classes becoming similar increases. This ambiguity leads to variation in classification, which increases the likelihood of missing pertinent information.

A crowd-sourced classification scheme should be able to deal with ambiguity (Malone, 2009; Sarasua, 2012). We propose to introduce *fuzzy* faceted information. No major classification scheme has used fuzzy information sets before (Kang, 2005; Baruchelli 2013; Absalom, 2012). Whereas Boolean faceted information retrieves a set of documents, fuzzy faceted information retrieves a ranked list. This can make allowance for subjectivity in fuzzy values and provide a degree of tolerance to inaccuracies in classification. Navigation to the correct classes in each facet is via a hyperlinked library of pictures of each individual class together with similar classes. The pictures are effectively one-dimensional ontologies of holistic similarity. Related entities appear above the class, the higher above the less like the class they are. Specific instances of the class appear underneath the class. In order for navigation to be both simple and intuitive, each class has an accompanying pop-up definition. At the same time, the pictures are the heart of the search engine.

We know of no classification scheme that implements *fuzzy faceted* classification on the basis of similarity between multiple entities. Prior art search includes the problem in the query. A full hit in stent, surface-materials, use in situ, and occlusion may return existing solutions such as a polished interior surface or drug-eluding stent techniques.

---

[3] http://www.cooperativepatentclassification.org/index.html;jsessionid=nvjip89ahf09
[4] Lofti A. Zadeh, stated: "as complexity rises, precise statements lose meaning and meaningful statements lose precision"

For a solution search the solution (obviously) is removed from the query. A search for a solution to the stent fouling problem may return a broad range of solutions, given that they are not specified. A catheter, for example, is similar to a stent. A stent has the problem of the build-up of bacteria, which is related to the problem of catheter occlusion. To remedy this, catheters use a lithographed surface, which is similar to a sharkskin surface. The lithographed surface may be the inspiration for non-smooth surfaces: like the dimples on a golf ball, the humpback whale like lumps on a helicopter blade, or the sharkskin on the hull of a ship.

An occasional searcher need simply identify a class in each of the facets. The system relies on experts, or teams of experts, assuming responsibility for the design of the pictures. Experts are free to pick and place classes from a central registry where a controlled vocabulary has a controlled meaning. Each class in the registry need not be mutually exclusive, but it does need to be distinct and unambiguous. Whilst there is a single meaning across the system, different experts can judge the significance in conferring a bespoke degree of similarity.

We believe that experts will be motivated to design the graphical representations for no more than the opportunity of making a contribution: exercising their expertise for the pleasure, or the 'love and glory' (Malone 2009). Importantly experts from around the world can translate the class definitions and classify and search in their mother tongues. That experts are free to design pictures from their perspective enables distributed decision-making. The combination of the different perspectives is performed by linking them in a 'fractal search' where each class on a picture is considered as an ontology rather than a single class. The system of fuzzy similarity allows the federation of existing information with a simple pick-and-place (Sarasua, 2012). Proprietary schemes (Wolter, 2012) and different patent schemes could be federated into a single system, after which a searcher would be no longer faced with a complex mosaic of disparate classification and indexing sources.

## 4. CONCLUSIONS AND FUTURE WORK

We propose a 5-faceted fuzzy classification scheme as the heart of a crowd-sourced initiative to classify patent literature and academic non-patent science and technology literature. The result will be a fractally-structured pior art repository which can be utilized to provide answers to multi-dimensional queries in two fields of application: (1) a search for prior art in new patent applications submitted to patent offices; (2) an inventive solution search for specific problems that may already have been solved in other domains.

Apart from organizing the consortium to get this initiative started, we seek the support of the international patent organizations to the make their patent databases available for this effort. Our design can also help academic publishers for classifying their collective science and technology archives in a novel way, allowing wider access and stimulating innovation.

The authors thank Elena Simperl and Lora Aroyo for feedback on earlier versions and helpful discussions.